# Asset Prices and Risk Aversion

*Abstract*

The standard asset pricing models (the CCAPM and the Epstein-Zin non-expected utility model) counterintuitively predict that equilibrium asset prices can rise if the representative agent's risk aversion increases. If the income effect, which implies enhanced saving as a result of an increase in risk aversion, dominates the substitution effect, which causes the representative agent to reallocate his portfolio in favour of riskless assets, the demand for securities increases. Thus, asset prices are forced to rise when the representative agent is more risk adverse. By disentangling risk aversion and intertemporal substituability, we demonstrate that the risky asset price is an increasing function of the coefficient of risk aversion only if the elasticity of intertemporal substitution (EIS) exceeds unity. This result, which was first proved par Epstein (1988) in a stationary economy setting with a constant risk aversion, is shown to hold true for non-stationary economies with a variable or constant risk aversion coefficient. The conclusion is that the EIS probably exceeds unity.

## Introduction

What is the influence of a risk aversion variation on risky asset prices? Intuitively, an increase (decrease) in risk aversion induces a fall (a rise) in securities prices. The more risk averse the representative agent is, the more that agent will prefer secure investments. As a result, his demand for risky assets decreases, pushing asset prices to fall[1]. However, according to the standard asset pricing models (the CCAPM and the Esptein-Zin's generalised version), an increase in risk aversion may induce risky asset prices to rise. To prevent this eventuality, restrictions on preference parameters must be imposed. We demonstrate that the CCAPM, with preferences of the standard Von Neumann-Morgenstern time-separable form, assesses an *a priori* unacceptable prediction, whereas the asset pricing model, derived by Epstein and Zin (1989, 1991) and Weil (1989, 1990) assumes a more general recursive utility specification and offers an acceptable prediction if the elasticity of intertemporal substitution (EIS) exceeds unity[2]. This study provides theoretical evidence in favour of a high value of the EIS, contrasting with certain empirical evidence that is documented in the literature (Hall, 1988, Ogaki and Reinhart, 1998, Campbell, 2003, Yogo, 2004, Braun and Nakajima, 2012).

Section 1 describes the environment and utility functions. Section 2 derives the representative agent's consumption-portfolio choice and solves for equilibrium prices, risk premium and the risk free rate. A static comparative analysis is conducted in Section 3 to study stock price implications of risk aversion increases. In Section 4, we consider a generalised version of the model, which allows the risk aversion coefficient to vary. A comparative-dynamics analysis is conducted to study the implications of variations in risk aversion. The final section presents the conclusions.

## 1. The utility functions and the environment

The environment is similar to the one described by Lucas (1978), Mehra and Prescott (1985), Epstein (1988), Weil (1989) and Epstein and Zin (1991). A perishable consumption good, a fruit, is produced by non-reproducible identical trees whose number is normalised to one, without loss of generality. Let $q_t$ denote the dividend (the number of fruits falling from the tree) collected at time t, associated with holding the single equity share. It is assumed that the production growth, $y_{t+1} = q_{t+1}/q_t$, follows an i.i.d. lognormal process :

$$\ln y_{t+1} = \Delta \ln q_{t+1} \sim \text{i.i.d.}\ N(\mu, \sigma^2). \quad (1)$$

Let $p_t$, $x_{t-1}$, $c_t$ and $b_{t-1}$ denote, respectively, the price at time t of the equity share (prices are in terms of the time t consumption good), the number of shares held by the representative agent at the beginning of period t (before trading starts), his consumption at t, and the number of the one-period real bill that the representative holds at the beginning of period t, which pays at time t one unit of the consumption good. The price of this riskless security is $1/R_F$, where $R_F - 1$ denotes the risk free rate. The budget constraint faced by the representative consumer is then:

$$c_t + p_t x_t + (b_t / R_F) = (p_t + y_t) x_{t-1} + b_{t-1}. \quad (2)$$

Let $R_{t+1} = (p_{t+1} + y_{t+1})/p_t$ denote the equity's one-period gross return and $w_t = (p_t + y_t)x_{t-1} + b_{t-1}$ represent the beginning-of-period wealth. The capital can either be consumed or invested in a portfolio combining risky and riskless assets. The representative consumer invests fraction $v_t = p_t x_t /(p_t x_t + b_t / R_F)$ of the non-consumed part of capital $w_t - c_t$ in the risky asset and the remaining fraction $1 - v_t = (b_t / R_F)/(p_t x_t + b_t / R_F)$ in the riskless asset. The budget constraint (2) can be rewritten more compactly as:

$$w_{t+1} = R_{pt+1}(w_t - c_t), \quad (3)$$

where $R_{pt+1}$ is the gross return of the representative agent's portfolio:

$$R_{pt+1} = v_t R_{t+1} + (1 - v_t) R_F. \quad (4)$$

---

[1] The financial markets' history confirms this rule. Because asset prices fall during stock market panics, we can consider them periods of increasing risk aversion.

[2] Epstein (1988) first proves this result in a stationary economy model by means of a comparative statics risk aversion analysis. We extend this result to the case of a more general non-stationary economy model by means of a comparative static or dynamics analysis.

We consider an infinitely-lived representative agent who receives utility from consumption of the goods in each period. In any period t, current consumption $c_t$ is non-stochastic but future consumption is generally uncertain. There are two key assumptions underlying the specification of utility. For the agent making a decision in period t, utility $U_{t+1}$ from period t+1 onward is random. First, we assume that the agent computes, given the information available to him, a certainty equivalent $\mu(U_{t+1}/I_t)$ of random future utility. Second, we assume that to obtain current-period lifetime utility, the agent combines $\mu(U_{t+1}/I_t)$ with current consumption via an aggregator function W[.,.], so that lifetime utility is given by:

$$U_t = W[c_t, \mu(U_{t+1}/I_t)]. \quad (5)$$

Consistent with Epstein and Zin (1989) and Weil (1990), this class of preferences that builds on the recursive structure of Kreps and Porteus (1978) allows a separation of risk aversion from intertemporal substitution that is not possible in the expected utility framework. We follow Epstein and Zin (1991) to specify W as the CES form:

$$W(c, z) = \left[c^{1-\rho} + \beta z^{1-\rho}\right]^{1/(1-\rho)}, \quad 1 \neq \rho > 0,$$

$$W(c, z) = \log(c) + \beta \log(z), \quad \rho = 1,$$

where $\rho$ is interpreted as an intertemporal fluctuations aversion parameter (the inverse of the IES), and $0 \leq \beta \leq 1$ is the subjective discount factor.

We adopt, as in Epstein and Zin (1991), a $\gamma$-mean (or constant relative risk aversion expected utility) specification for the certainty equivalent:

$$\mu[x] = \left(Ex^{1-\gamma}\right)^{1/(1-\gamma)}, \quad 1 \neq \gamma > 0,$$

$$\log(\mu) = E[\log(x)], \quad \gamma = 1,$$

where E is the expectation operator, and where $\gamma$ may be interpreted as a relative risk aversion parameter.

This leads to the recursive structure for intertemporal utility (if $\rho \neq 1$ and $\gamma \neq 1$):

$$U_t = \left[c_t^{1-\rho} + \beta \left(E_t U_{t+1}^{1-\gamma}\right)^{(1-\rho)/(1-\gamma)}\right]^{1/(1-\rho)} \quad (6),$$

where $E_t$ is the conditional expectation operator given $I_t$.

When $\gamma = \rho$, (6) specialise to the common expected utility specification:

$$U_t = \left[E_t \sum_{i=0}^{\infty} \beta^i c_{t+i}^{1-\gamma}\right]^{1/(1-\gamma)}. \quad (7)$$

## 2. Optimal consumption, portfolio choice and equilibrium

Given the recursive structure (5), intertemporal utility maximisation leads to the following dynamic programming problem:

$$J(w_t) = \max_{c_t, b_t} W[c_t, \mu\{J(w_{t+1})/I_t\}], \quad (8)$$

where $J(w_t)$ denotes the maximum utility achievable given the beginning-of-period wealth $w_t$.

Epstein (1988), Weil (1989), Giovannini and Weil (1989) and Epstein et Zin (1989) solved this dynamic problem, and attained the following Euler equations (if $\rho \neq 1$ and $\gamma \neq 1$):

$$E_t\left\{\left[\beta\left(\frac{c_{t+1}}{c_t}\right)^{-\rho}\right]^{\frac{(1-\gamma)}{(1-\rho)}}[R_{pt+1}]^{\frac{(1-\gamma)}{(1-\rho)}-1}R_{t+1}\right\} = 1, \quad (9a)$$

$$E_t\left\{\left[\beta\left(\frac{c_{t+1}}{c_t}\right)^{-\rho}\right]^{\frac{(1-\gamma)}{(1-\rho)}}[R_{pt+1}]^{\frac{(1-\gamma)}{(1-\rho)}-1}R_F\right\} = 1. \quad (9b)$$

When $\rho = \gamma$, (9a) and (9b) specialise to the familiar C-CAPM's Euler equations (Rubinstein, 1976 and Lucas, 1978).

In equilibrium, the entirety of period t's perishable output is consumed during that period: $c_t = q_t \; \forall t$, and the bond market is cleared: $b_t = 0 \; \forall t$. The representative agent's portfolio is then composed completely of risky assets: $v_t = 1, R_{pt+1} = R_{t+1}$, and the Euler equations (9a) et (9b) simplify to:

$$E_t\left\{\left[\beta y_{t+1}^{-\rho}\right]^{\frac{(1-\gamma)}{(1-\rho)}}[R_{t+1}]^{\frac{(1-\gamma)}{(1-\rho)}}\right\} = 1, \quad (10a)$$

$$E_t\left\{\left[\beta y_{t+1}^{-\rho}\right]^{\frac{(1-\gamma)}{(1-\rho)}}[R_{t+1}]^{\frac{(1-\gamma)}{(1-\rho)}-1}R_F\right\} = 1. \quad (10b)$$

It can be demonstrated that the homogeneous price function $p_t = cq_t$, with c > 0, solves equation (10a). We demonstrate in the appendix that c satisfies to the following condition[3]:

$$\left(\frac{c}{1+c}\right) = \exp\left[-\delta + (1-\rho)\mu + \frac{1}{2}(1-\rho)(1-\gamma)\sigma^2\right], \text{ with } \beta = e^{-\delta}. \quad (11)$$

In equilibrium, the risk free rate and the risk premium are:

$$\ln R_F = \delta + \rho\left(\mu + \frac{1}{2}\sigma^2\right) - \frac{1}{2}\gamma(1+\rho)\sigma^2, \quad (12)$$

$$\ln E(R_{t+1}) - \ln R_F = \gamma\sigma^2. \quad (13)$$

The risk free rate is high when the discount parameter $\delta$ is high; a high interest rate is required to convince investors to save rather than to consume. The risk free rate is also high when the logarithmic expected growth rate[4] $\mu + \sigma^2/2$ and the intertemporal fluctuations aversion parameter $\rho$ are high. As the representative consumer feels aversion for intertemporal fluctuations, the desire is to consume more when the expected growth rate is positive. The real interest rate is then pushed upward to restore equilibrium. Finally, the risk free rate is high when volatility of consumption $\sigma^2$ is low. Volatility of consumption $\sigma^2$ captures precautionary saving; the representative consumer is more concerned with low consumption states than he is pleased by high consumption states.

---

[3] See the Appendix for a derivation of equations (11), (12), (13), (14), (20a) and (20b).
[4] According to (1), the production growth rate follows a lognormal process, therefore, $\ln E(y_{t+1}) = \mu + \sigma^2/2$.

Equation (13) implies that the equity premium is the product of the coefficient of risk aversion and the variance of the growth rate of consumption.

## 3. A static comparative analysis

Epstein (1988) demonstrates, assuming a stationary environment[5] that the effect on risky asset prices of a variation of the risk aversion parameter depends on the value of the intertemporal elasticity of substitution. We demonstrate that this result holds true if the economy is non-stationary.

Solve for c in equation (11) and substitute in the homogenous price function $p_t = c q_t$ to deduce that:

$$p_t = \frac{\exp\left[\mu - \frac{\sigma^2}{2} - \ln E(R_{t+1})\right] q_t}{1 - \exp\left[\mu - \frac{\sigma^2}{2} - \ln E(R_{t+1})\right]}. \quad (14)$$

The derivative $\partial p_t / \partial \gamma$ has the opposite sign to that of $\partial \ln E(R_{t+1}) / \partial \gamma$, namely:

$$\frac{\partial \ln E(R_{t+1})}{\partial \gamma} = \frac{\partial \ln R_F}{\partial \gamma} + \frac{\partial (\ln E(R_{t+1}) - \ln R_F)}{\partial \gamma}. \quad (15)$$

The consequences of increased risk aversion on equity returns are ambiguous because the risk premium will move in the opposite direction to interest rate changes according to equations (12) and (13). Then, if $\gamma$ rises, the risk free rate may fall sufficiently to induce a drop in expected returns, despite an increase in the risk premium.

When $\gamma = \rho$ in the familiar CCAPM environment, equation (15) can be written as:

$$\frac{\partial \ln E(R_{t+1})}{\partial \gamma} = \mu + \sigma^2(1 - \gamma), \quad \text{for } \gamma = \rho. \quad (16)$$

Therefore, the expected return on equity is not a monotonic function of $\gamma$ as Donaldson and Mehra (1984) first observed by means of numerical simulations. For any stationary environment ($\mu = 0$), this derivative is negative when the risk aversion coefficient is higher than one, which is highly likely. For any growth economy ($\mu > 0$), the derivative is negative or positive depending on whether the growth rate is sufficiently high enough. Thus it appears that the CCAPM is not suitable for modelling a low-growth economy: an increased risk aversion induces risky asset prices to rise.

For general parameter values, when the coefficient of risk aversion and the elasticity of substitution are clearly separated, equations (12) and (13) imply:

$$\frac{\partial \ln E(R_{t+1})}{\partial \gamma} = \frac{\sigma^2}{2}(1 - \rho). \quad (17)$$

This result is similar to that achieved by Epstein (1988) and can be interpreted in the same way. An increase in risk aversion acts to reduce the certainty equivalent return to saving. If $\rho < (>)1$, the dominant income (substitution) effect implies reduced (enhanced) present consumption and an increased (reduced) demand for securities. Thus, expected equity return is forced to decrease (to increase).

---

[5] Letting $\mu = 0$ in our model leaves a stationary economy with an i.i.d. production growth rate $y_t$, whereas Epstein (1988) considers a stationary economy where it is the production level $q_t$ that is i.i.d.

## 4. A dynamic comparative analysis

We demonstrate in this section that dynamic and static comparative analyses provide similar results. Assume that the intertemporal utility is of the following form:

$$U_t = W(c_t, E_t U_{t+1}) = \left\{ c_t^{1-\rho} + \beta (E_t U_{t+1})^{\frac{1-\rho}{1-\gamma_t}} \right\}^{\frac{1-\gamma_t}{1-\rho}} \quad (18)$$

which is of the same form as (6), except that $\gamma_t$ substitutes for $\gamma$.

The utility function (18) is characterised by a time-variable risk aversion parameter $\gamma_t$. Let us suppose that the variations of $\gamma_t$ are non-stochastic (completely predictable). The Bellman equation takes the form:

$$J(w_t, t) = \max_{c_t, b_t} \left[ c_t^{(1-\rho)} + \beta E_t^{(1-\rho)/(1-\gamma_t)} \left[ J^{(1-\gamma_t)}(w_{t+1}, t+1) \right] \right]^{1/(1-\rho)}. \quad (19)$$

First-order conditions for this maximisation problem lead to the following equations:

$$\beta^{\frac{1-\gamma_t}{1-\rho}} E_t \left[ y_{t+1}^{-\frac{\rho(1-\gamma_t)}{(1-\rho)}} R_{t+1}^{\frac{1-\gamma_t}{1-\rho}} \right] = 1, \quad (20a)$$

$$\beta^{\frac{1-\gamma_t}{1-\rho}} E_t \left[ y_{t+1}^{-\frac{\rho(1-\gamma_t)}{(1-\rho)}} R_{t+1}^{\frac{1-\gamma_t}{1-\rho}-1} R_{Ft+1} \right] = 1. \quad (20b)$$

Equations (20a) and (20b) are similar to equations (10a) and (10b), except that $\gamma_t$ substitutes for $\gamma$. The expressions for equilibrium risk free rate and risk premium are obtained by substituting $\gamma_t$ for $\gamma$ in equations (12) and (13).

It can be demonstrated that the homogeneous price function $p_t = c_t q_t$ solves equation (20a), where $c_t > 0$ obeys the following difference equation:

$$\frac{c_t}{1 + c_{t+1}} = \exp\left[-\delta + (1-\rho)\mu + (1-\rho)(1-\gamma_t)\sigma^2/2\right]. \quad (21)$$

Because $p_{t+1}/p_t = (c_{t+1}/c_t)(q_{t+1}/q_t)$, the asset price change induced by a variation of $\gamma_t$ depends only on $c_{t+1}/c_t$. Let $h_t$ denote the right-hand side of equation (21). The difference equation (21) can be solved forward:

$$c_t = \sum_{n=0}^{\infty} \prod_{j=0}^{n} h_{t+j} = h_t + h_t h_{t+1} + h_t h_{t+1} h_{t+2} + h_t h_{t+1} h_{t+2} h_{t+3} + \ldots \quad (22)$$

If the risk aversion parameter is to rise permanently by $\Delta > 0$, so that $\gamma_{t+j} = \gamma_t + \Delta \; \forall j > 0$, it therefore follows that:

$$h_{t+j} = h_{t+1} \; \forall j > 0, \text{ where } h_{t+1} = h_t \exp\left[-\Delta(1-\rho)\sigma^2/2\right]. \quad (23)$$

Under such conditions, current and next values of $c_t$ are:

$$c_t = h_t + h_t h_{t+1} + h_t h_{t+1} h_{t+1} + h_t h_{t+1} h_{t+1} h_{t+1} + \ldots, \quad (24a)$$

$$c_{t+1} = h_{t+1} + h_{t+1} h_{t+1} + h_{t+1} h_{t+1} h_{t+1} + h_{t+1} h_{t+1} h_{t+1} h_{t+1} + \ldots \quad (24b)$$

From these equations, we can deduce the following equivalence:

$$\frac{c_{t+1}}{c_t} < 1 \Leftrightarrow \frac{h_{t+1}}{h_t} = \exp\left[-\Delta(1-\rho)\sigma^2/2\right] < 1. \quad (25)$$

A permanent risk aversion increase results in a risky asset price fall only if $\rho < 1$. This confirms the results of the static comparative analysis.

If the risk aversion parameter is to rise temporarily by $\Delta > 0$, so that $\gamma_{t+1} = \gamma_t + \Delta$ and $\gamma_{t+j} = \gamma_t \ \forall j > 1$, it follows that:

$$h_{t+1} = h_t \exp\left[-\Delta(1-\rho)\sigma^2/2\right], \text{ and } h_{t+j} = h_t \ \forall j > 1. \quad (26)$$

We therefore deduce from (22) that:

$$c_t = h_t + h_t h_{t+1} + h_t h_{t+1} h_t + h_t h_{t+1} h_t h_t + ..., \quad (27a)$$

$$c_{t+1} = h_{t+1} + h_{t+1} h_t + h_{t+1} h_t h_t + h_{t+1} h_t h_t h_t + ... \quad (27b)$$

The right-hand sides of these two equations differ only by the first term. Therefore, the equivalence (25) holds true for a temporary risk aversion increase. Thus, a transitory risk aversion increase results in a risky asset price fall only if $\rho < 1$.

## 5. Conclusions

Standard asset pricing models predict that a risk aversion increase can cause risky asset prices to rise. If researchers in asset pricing modeling reject such an unrealistic prediction, they should exercise caution with the preference specification of their models. In particular, the common expected utility model should be set aside in favour of a constrained version of the Epstein-Zin (1989, 1991) and Weil (1989, 1990) recursive utility model, which is characterised by an intertemporal fluctuations aversion parameter (the inverse of the EIS) of less than one. This result contrasts with the direct empirical evidence for the low elasticity of the intertemporal substitution in consumption that has been documented by many researchers (Hall, 1988, Ogaki and Reinhart, 1998, Campbell, 2003, Yogo, 2004, Braun and Nakajima, 2012), and confirms our belief that EIS probably exceeds unity.

# Appendix

**The derivation of equation (11):**

Substituting $p_t = cq_t$ and $p_{t+1} = cq_{t+1}$ in (10a), we find that $\left(\dfrac{c}{1+c}\right) = \beta \left[E\left\{y_{t+1}^{(1-\gamma)}\right\}\right]^{\frac{(1-\rho)}{(1-\gamma)}}$. Using the property of a normal variable z: $E[e^z] = e^{E[z] + \frac{1}{2}V[z]}$, this last equation can be rearranged to obtain equation (11).

**The derivation of equation (12):**

Observe that the gross return on equity is proportional to the lognormal dividend growth rate:
$R_{t+1} = \dfrac{p_{t+1} + q_{t+1}}{p_t} = \dfrac{cq_{t+1} + q_{t+1}}{cq_t} = \dfrac{1+c}{c} \dfrac{q_{t+1}}{q_t} = \dfrac{1+c}{c} y_{t+1}$. Then, $(\ln R_{t+1}, \ln y_{t+1})$ are jointly normally distributed. Moreover, because the growth rate of dividend and the return on equity are i.i.d., the conditional and unconditional expectations of any function of $y_{t+1}$ and $R_{t+1}$ are the same. Then, equation (10b) can be written in the form:

$$\beta^{\frac{(1-\gamma)}{(1-\rho)}} \left(\dfrac{1+c}{c}\right)^{\frac{(1-\gamma)}{(1-\rho)}-1} E\left\{y_{t+1}^{-\gamma}\right\} R_F = 1 .$$

Substituting for (1+c)/c from equation (11) and using the lognormal distribution assumption, we obtain:

$$\exp\left\{-\delta\dfrac{(1-\gamma)}{(1-\rho)}\right\}\left[\exp\left\{-\delta + (1-\rho)\mu + \dfrac{1}{2}\dfrac{(1-\rho)}{(1-\gamma)}(1-\gamma)^2\sigma^2\right\}\right]^{1-\frac{(1-\gamma)}{(1-\rho)}} \exp\left\{-\gamma\mu + \dfrac{1}{2}\gamma^2\sigma^2\right\} R_F = 1.$$

Taking logs on both sides and simplifying we obtain equation (12).

**The derivation of equation (13):**

Given that $(\ln R_{t+1}, \ln y_{t+1})$ are jointly normally distributed, equation (10a) can be rearranged:

$$\dfrac{(1-\gamma)}{(1-\rho)}[\ln \beta - \rho\mu] + \dfrac{(1-\gamma)}{(1-\rho)} E(\ln R_{t+1}) + \dfrac{1}{2}\dfrac{(1-\gamma)^2}{(1-\rho)^2} V(\ln R_{t+1} - \rho \ln y_{t+1}) = 0$$

Note that $\ln R_{t+1} = \ln\dfrac{1+c}{c} + \ln y_{t+1}$ to obtain:

$$E(\ln R_{t+1}) = \delta + \rho\mu - \dfrac{1}{2}(1-\rho)(1-\gamma)\sigma^2 \quad (A1)$$

Subtracting (12) from (A1), we find that:

$$E(\ln R_{t+1}) + \dfrac{1}{2}\sigma^2 - \ln R_F = \gamma\sigma^2 \qquad (A2)$$

The lognormal distribution assumption implies that:

$$\ln E(R_{t+1}) = E(\ln R_{t+1}) + \dfrac{1}{2}V(\ln R_{t+1}), \qquad (A3)$$

Substituting (A3) in (A2) results in $\ln E(R_{t+1}) - \ln R_F = \gamma\sigma^2$.

**The derivation of equation (14):**

Solve equation (11) for c and substitute in $p_t = cq_t$ to obtain:

$$p_t = \frac{\exp\left[-\delta + (1-\rho)\mu + \frac{1}{2}(1-\rho)(1-\gamma)\sigma^2\right] q_t}{1 - \exp\left[-\delta + (1-\rho)\mu + \frac{1}{2}(1-\rho)(1-\gamma)\sigma^2\right]}. \quad (A4)$$

Combine (A1) and (A3) to find equation (14).

**The derivation of equations (20a) and (20b):**
Maximisation of (18) leads to the dynamic programming problem $J(w_t, t) = \max_{c_t, b_t} W[c_t, \mu\{J(w_{t+1}, t+1)/I_t\}]$, where $J(w_t, t)$ denotes the maximum utility achievable in period t, given the beginning-of-period wealth $w_t$. The homogeneity of $U_t$ and the linearity of $w_{t+1}$ in $(w_t, c_t)$, by equation (3), imply that the value function has the form $J(w_t, t) = A(t) w_t$, $A(t) > 0$. Then, the Bellman equation can be written:

$$A(t)w_t = \max_{c_t, b_t}\left[c_t^{1-\rho} + \beta(w_t - c_t)^{1-\rho} E_t^{(1-\rho)/(1-\gamma_t)}\left\{[A(t+1)R_{pt+1}]^{(1-\gamma_t)}\right\}\right]^{1/(1-\rho)}. \quad (A5)$$

This maximisation problem can be decomposed into two maximisation problems. Portfolio choice can be described as:

$$\mu_t^* = \max_{b_t} E_t^{1/(1-\gamma_t)}\left\{[A(t+1)\{b_t R_{t+1} + (1-b_t)R_{Ft+1}\}]^{(1-\gamma_t)}\right\}, \quad (A6)$$

and consumption is chosen by:

$$A(t)w_t = \max_{c_t}\left[c_t^{(1-\rho)} + \beta(w_t - c_t)^{(1-\rho)}\mu_t^{*(1-\rho)}\right]^{1/(1-\rho)}. \quad (A7)$$

The homogeneity of (A7) implies that the optimal consumption can be written $c_t = a(t)w_t$. Substituting this expression into (A7), we obtain:

$$A^{(1-\rho)}(t) = a^{(1-\rho)}(t) + \beta(1-a(t))^{(1-\rho)}\mu_t^{*(1-\rho)}. \quad (A8)$$

The first-order condition in (A7) implies:

$$a^{-\rho}(t) = \beta(1-a(t))^{-\rho}\mu_t^{*(1-\rho)}. \quad (A9)$$

These last two equations combine to yield $A(t) = [a(t)]^{-\frac{\rho}{(1-\rho)}} = (c_t/w_t)^{-\frac{\rho}{(1-\rho)}}$, thus:

$$A(t+1) = (c_{t+1}/w_{t+1})^{-\frac{\rho}{(1-\rho)}} = (c_{t+1}/R_{pt+1})^{-\frac{\rho}{(1-\rho)}}(w_t - c_t)^{\frac{\rho}{(1-\rho)}}. \quad (A10)$$

Substituting (A6) and (A10) into (A9) provides:

$$\beta^{(1-\gamma_t)/(1-\rho)} E_t\left[(c_{t+1}/c_t)^{-(1-\gamma_t)\rho/(1-\rho)} R_{pt+1}^{(1-\gamma_t)/(1-\rho)}\right] = 1. \quad (A11)$$

Substituting (A10) into (A6), the portfolio choice problem becomes $\mu_t^* = \max_{b_t}(w_t - c_t)^{\frac{\rho}{(1-\rho)}} E_t^{1/(1-\gamma_t)}\left[c_{t+1}^{-\frac{(1-\gamma_t)\rho}{(1-\rho)}} R_{pt+1}^{\frac{(1-\gamma_t)}{(1-\rho)}}\right]$. Maximise this last equation to obtain:

$$\beta^{(1-\gamma_t)/(1-\rho)} E_t\left[\left(\frac{c_{t+1}}{c_t}\right)^{-\frac{(1-\gamma_t)\rho}{(1-\rho)}} R_{pt+1}^{\frac{(1-\gamma_t)}{(1-\rho)}-1}(R_{t+1} - R_{Ft+1})\right] = 0. \quad (A12)$$

Substitute equilibrium conditions $y_{t+1} = c_{t+1}/c_t$ and $R_{pt+1} = R_{t+1}$ into (A11) to get equation (20a):

$$\beta^{\frac{1-\gamma_t}{1-\rho}} E_t\left[y_{t+1}^{-\frac{\rho(1-\gamma_t)}{(1-\rho)}} R_{t+1}^{\frac{1-\gamma_t}{1-\rho}}\right] = 1. \quad (20a)$$

Then, substitute (20a) and equilibrium conditions into (A12) to obtain equation (20b):

$$\beta^{\frac{1-\gamma_t}{1-\rho}} E_t \left[ y_{t+1}^{-\frac{\rho(1-\gamma_t)}{(1-\rho)}} R_{t+1}^{\frac{1-\gamma_t}{1-\rho}-1} R_{Ft+1} \right] = 1 \ . \quad (20b)$$